\documentclass[journal]{IEEEtran}

\usepackage{latexsym}
\usepackage{amsfonts}
\usepackage{amsbsy}
\usepackage{amsmath,amssymb}
\usepackage{times}
\usepackage{graphicx}
\usepackage{enumerate}
\usepackage[usenames]{color}
\usepackage[dvips]{pstcol}
\usepackage{epstopdf}
\usepackage{amsmath}
\usepackage{subfig}
\usepackage{bm}
\usepackage{booktabs}
\usepackage{cite}
\usepackage{color}
\usepackage{xcolor}
\usepackage{algorithm}
\usepackage{algorithmicx}
\usepackage{algpseudocode}
\usepackage{setspace}
\usepackage{bm}
\usepackage{tabularx}
\usepackage{amstext,epsfig,psfrag,graphicx,cite,footnote}
\usepackage{siunitx}
\usepackage{mathrsfs}
\usepackage{booktabs}
\usepackage{multirow}
\usepackage{stfloats}
\ifCLASSINFOpdf
\else
\fi
\begin{document}
\title{A PDD Decoder for Binary Linear Codes With Neural Check Polytope Projection}
% note positions of commas and nonbreaking spaces ( ~ ) LaTeX will not breaki
\author{Yi Wei, Ming-Min Zhao, Min-Jian Zhao and Ming Lei\vspace{-2.5em}
\thanks{Y. Wei, M. M. Zhao, M. J. Zhao and M. Lei are with College of Information Science and Electronic Engineering, Zhejiang University, Hangzhou 310027, China
(email: \{21731133, zmmblack, mjzhao, lm1029\}@zju.edu.cn) (Corresponding author: Ming-Min Zhao and Min-Jian Zhao.)

This work was supported in part by  the
National Key R$\&$D Program of China under Grant 2018YFB1802303, in part by the National Natural Science Foundation of China under Grant 91938202, in part by the Zhejiang Provincial Natural Science Foundation of China under Grant LQ20F010010, in part by the Fundamental Research Funds for the Central Universities under Grant 2019QNA5011.
}}
\maketitle


\begin{abstract}
\color{black}{
Linear Programming (LP) is an important decoding technique for binary linear codes. However, the advantages of LP decoding, such as low error floor and strong  theoretical guarantee, etc., come at the cost of high computational complexity and poor performance at the low signal-to-noise ratio (SNR) region. In this letter, we  adopt the penalty dual decomposition (PDD) framework and propose a PDD algorithm to address the fundamental polytope based maximum likelihood (ML) decoding problem. Furthermore, we propose to integrate machine learning techniques into the most time-consuming part of the PDD decoding algorithm, i.e., check polytope projection (CPP). Inspired by the fact that a multi-layer perception (MLP) can theoretically approximate any nonlinear mapping function, we present a specially designed neural CPP (NCPP) algorithm to decrease the decoding latency. Simulation results demonstrate the effectiveness of the proposed algorithms.
}
\end{abstract}

\begin{IEEEkeywords}
Binary linear codes, check polytope projection, LDPC, machine learning, MLP, neural network.
\end{IEEEkeywords}

% \fi
\IEEEpeerreviewmaketitle

\vspace{-0.3em}
\section{Introduction}
Recently, the linear programming (LP) decoder, which is based on LP relaxation of the maximum likelihood (ML) decoding problem, has attracted increasing attention for decoding binary linear codes, especially for low density parity check (LDPC) codes \cite{1397933,6187728}. Compared with the classical belief propagation (BP) decoder, the LP decoder has stronger theoretical guarantees on decoding performance and empirically is not observed to suffer from an error floor. However, the above advantages of the LP decoder come at the cost of two drawbacks, i.e., higher computational complexity and poorer error-correcting performance when the signal-to-noise-ratio (SNR) is low.
%in the low  signal-to-noise-ratio (SNR) region compared with the classical belief propagation (BP) decoder.
%in the low  signal-to-noise-ratio (SNR) region compared with the classical belief propagation (BP) decoder.

 In order to overcome the above shortcomings, the work \cite{LiuIT2013} first employed the alternating direction method
of multipliers (ADMM) to solve the ML decoding problem by exploiting the fundamental polytope of the parity-check (PC) constraints. In order to improve the error-rate performance in the low SNR
region, the work \cite{7456284} added penalty terms to the linear objective function to make pseudocodewords more costly and the resulting decoder is known as the ADMM penalized decoder. Afterwards, improvements over the ADMM penalized decoder were achieved by modifying the penalty terms \cite{7740920}, \cite{JiaoCL2015}.
%the penalty terms were improved ADMM-penalized decoder was further improved by using piecewise penalty functions, and for irregular LDPC codes, the work \cite{JiaoCL2015} proposed to  modify the penalty term of the objective function and assign different penalty parameters for variable nodes with different degrees.
{ Based on the  cascaded decomposition
method \cite{4455768}, %J. Bai et al. proposed a quadratic programming (QP) ADMM decoder with linear complexity and better performance compared to the BP decoder.
the work \cite{8691508} adopted the penalty dual decomposition (PDD) framework \cite{ShiPDD2017} and developed a PDD decoder, which was shown to overperform the ADMM penalized decoder.}
%can be implemented in fully parallel with linear
%complexity with a superior decoding performance.

Focusing on the fundamental polytope based ML decoding problem, intensive studies have been conducted to simplify  the check polytope projection (CPP) operations, which is the most computationally intensive and time-consuming part in the ADMM-based decoders \cite{ZhangISIT2013,6636134,Wei2018,7073564}.
%are dominated by the projection of a real-valued vector onto the
%parity-check (check) polytope.
Compared with the original CPP algorithm which involves two sorting operations \cite{LiuIT2013}, the work \cite{ZhangISIT2013} employed a cut-search algorithm to remove one sorting operation. In \cite{6636134}, the projection algorithm was further simplified by transforming the CPP operation into the projection onto   a simplex. In \cite{Wei2018}, the authors presented an iterative CPP (ICPP) algorithm which  requires no sorting operation and can substantially improve the decoding speed. The work \cite{ADMM1901}  revealed the recursive structure of the parity polytope and presented an efficient projection algorithm by iteratively fixing selected components of the projection.
%The work \cite{7073564} proposed to reduce the computational complexity of ADMM-based decoder
%by reducing the number of CPP rather than
%focussing on reducing the complexity of the projection itself.

%Moreover, recent years machine learning methods have demonstrated amazing performances in various tasks. As one of the most widely used methods in machine learning field, the multi-layer neural network, also called multi-layer perception (MLP), can theoretically map any continuous function if there are enough neurons.

In this letter, we propose a novel multi-layer perception (MLP)-aided PDD decoder for binary linear codes. Different from the PDD decoder in \cite{8691508} which considers the minimum polytope based LP formulation, we show that the PDD framework can also be applied to the fundamental polytope based LP formulation. The proposed decoder consists of two loops: in the outer loop, we update the dual variables and certain penalty parameter, while in the inner loop, we divide the primal variables into several blocks and employ the block coordinate descent (BCD) method to iteratively optimize each block variable in closed-form. Furthermore, in order to simplify the CPP operations in the proposed PDD decoder, we propose a neural CPP (NCPP) algorithm, which is obtained by integrating a simple three-layer MLP (namely CPP-net) into the ICPP algorithm in \cite{Wei2018} to reduce the corresponding iteration number. Simulation results demonstrate
that the proposed PDD decoder exhibits superior error-correcting performance and the NCPP algorithm can reduce the latency significantly.
\vspace{-0.3cm}
\section{Problem Formulation}
Consider a binary linear code $\mathcal{C}$ of length $N$ specified by an $M \times N$ PC matrix $\mathbf{H}$. Let $\mathcal{I} \triangleq \{1,\cdots,N\}$ and $\mathcal{J} \triangleq \{1,\cdots,M\}$ denote the sets of variables nodes and check nodes of $\mathcal{C}$, respectively.
%Let $d_j$ represents the degree of check node $j$.
Suppose $\mathbf{x}=\{x_i = \{0,1\},i \in \mathcal{I}\}$ is a codeword transmitted over a memoryless binary-input symmetric-output channels, and $\mathbf{y}$ is the received signal. Then, the received  log-likelihood ratio (LLR) vector $\mathbf{v} \in \mathbb{R}^{N\times 1}$ can be expressed as
\vspace{-0.2cm}
\begin{equation} \label{llr}
\small
v_i = \log \left( \frac{\textrm{Pr}(y_i | x_i = 0)}{\textrm{Pr}(y_i | x_i = 1)} \right),\;i \in \mathcal{I}.
\end{equation}

\vspace{-0.1cm}
\noindent According to \cite{7456284}, the ML decoding problem can be formulated as the following optimization problem:
\vspace{-0.13cm}
\begin{equation} \label{1}
\small
{\min \limits_{\mathbf{x}} \; \mathbf{v}^T\mathbf{x} \qquad \textrm{s.t.}\;  \mathbf{x} \in \mathcal{C}.}
\end{equation}
\vspace{-0.6cm}
%\noindent where $\mathcal{D}$ is the set of possible codewords.

\noindent The work \cite{1397933} proposed to relax problem \eqref{1} as follows:
\vspace{-0.13cm}
\begin{equation} \label{LP_problem}
\small
{\min \limits_{\mathbf{x}} \; \mathbf{v}^T\mathbf{x} \qquad \textrm{s.t.}\; \mathbf{x} \in \mathcal{P}:= \bigcap_{j\in\mathcal{J}} \textrm{conv}(\mathcal{D}_j),}
\end{equation}
\vspace{-0.3cm}

\noindent {where $\textrm{conv}(\mathcal{D}_j)$ is the convex hull of the codewords defined by the $j$-th row of the PC matrix $\mathbf{H}$, and $\mathcal{P}$ is called the fundamental polytope.} Let $d_j$ denote the degree of check node $j$, \eqref{LP_problem}  can be expressed in a more compact form as \cite{1397933}
\vspace{-0.10cm}
\begin{equation} \label{LP_problem_equi}
\small
{\min \limits_{\mathbf{x}} \; \mathbf{v}^T\mathbf{x} \qquad \textrm{s.t.}\; \mathbf{P}_j \mathbf{x} \in \mathbb{PP}_{d_j},\;\forall j \in \mathcal{J},}
\end{equation}

\vspace{-0.1cm}
\noindent where ${\mathbf{P}}_j$ denotes a $d_j\times N$ selection matrix which selects the elements of $\mathbf{x}$ that participate in the $j$-th check equation. ${\mathbb{PP}}_{d_j}$ is the PC polytope of dimension $d_j$, which is defined as the convex hull of all even-parity binary vectors of length $d_j$, i.e., ${\mathbb{PP}}_{d_j}=\textrm{conv}(\{{\mathbf{e}}\in\{0,1\}^{d_j} \vert \;\|\mathbf{e}\|_1 \textrm{ is even}\})$.
%${\mathbb{P}}_{d_j}$ is the set of even-weight binary vectors of length $d_j$. The parity-check polytope corresponding to the $j$-th check equation ${\mathbb{PP}}_{d_j}$ is defined as the the convex hull of the set ${\mathbb{P}}_{d_j}$.
%\section{NN aided PDD decoding}
%In this section,  we first adopt the PDD framework to solve the LP problem \eqref{LP_problem_equi} and develop a PDD decoding algorithm, which can achieve a performance gain over the ADMM-based decoders. Next, a novel Neural CPP algorithm is proposed to further reduce the computational complexity. This algorithm also can be applied to other LP decoders which exploit the fundamental polytope of PC constraints, e.g., some ADMM-based decoders.
\vspace{-0.2cm}
\section{PDD decoding algorithm}
\vspace{-0.0cm}
{ In this section,  we adopt the PDD framework to solve problem \eqref{LP_problem_equi} and develop a PDD decoding algorithm, where the main idea is to introduce additional equality constraints to handle the nontrivial constraints and the discrete variable $\mathbf{x}$.}
%Note that the difference between the proposed decoder and that in \cite{8691508} is that we employ the fundamental polytope based LP formulation in this work.
 %and since less auxiliary variables are needed in this case, the proposed PDD decoder exhibits better performance for longer code lengths (as will be shown in Section V).

Firstly, we introduce auxiliary variables
$\{{\mathbf{z}}_j\in {\mathbb{R}}^{d_j\times 1}\}$ to equivalently transform constraint $\mathbf{P}_j \mathbf{x} \in \mathbb{PP}_{d_j},\;\forall j \in \mathcal{J}$ into $ \mathbf{P}_j \mathbf{x} = \mathbf{z}_j, \mathbf{z}_j  \in \mathbb{PP}_{d_j},\;\forall j \in \mathcal{J}$.
%\begin{subequations} \label{LP_problem_PDD}
%	\begin{align}
%	&\min \limits_{\mathbf{x},\; \mathbf{z}} \; \mathbf{v}^T\mathbf{x}\\
%	&\textrm{s.t.}\; \mathbf{P}_j \mathbf{x} = \mathbf{z}_j, \mathbf{z}_j  \in \mathbb{PP}_{d_j},\;\forall j \in \mathcal{J}.
%	\end{align}
%\end{subequations}
Then, we relax the binary variables $\{x_i\}$ to the interval $[0,1]$, and instead of using penalty functions to enforce $x_i$ to $0$ or $1$, we propose to introduce auxiliary variables $\{\hat{x}_i\}$ which satisfy $\hat{x}_i =x_i$ and $x_i (\hat{x}_i-1) = 0$. Let $\hat{\mathbf{x}}=[{\hat{x}}_1,\cdots,{\hat{x}}_N]^T$ and $\mathbf{z}=[{\mathbf{z}}_1^T,\cdots,{\mathbf{z}}_M^T]^T$, problem \eqref{LP_problem_equi} can be equivalently formulated as
%In order to frame the decoding problem in the template of PDD, we can transform \eqref{LP_problem_equi} into the following problem:
\vspace{-0.13cm}
\begin{equation} \label{LP_problem_PDD}{
\small
	\begin{aligned}
	\min \limits_{\mathbf{x}, \;\hat{\mathbf{x}},\; {\mathbf{z}}} \; & \mathbf{v}^T\mathbf{x}\\
	\textrm{s.t.}  \;&\mathbf{P}_j \mathbf{x} = \mathbf{z}_j, \mathbf{z}_j  \in \mathbb{PP}_{d_j},\;\forall j \in \mathcal{J},\\
	& x_i(\hat{x}_i - 1) = 0,\; x_i = \hat{x}_i,\;0\leq x_i \leq 1,\;\forall i \in \mathcal{I}.
	\end{aligned} }
\end{equation}

\vspace{-0.2cm}
\noindent Next, we can see that the augmented Lagrangian problem of \eqref{LP_problem_PDD} can be expressed as {
\vspace{-0.13cm}
\begin{equation} \label{AL_problem}
\small
{ \begin{aligned}
\min \limits_{\mathbf{x},\;\hat{\mathbf{x}},\;\mathbf{z}} \; &\mathbf{v}^T\mathbf{x} + P_{\mu_m}(\mathbf{x},\hat{\mathbf{x}},\mathbf{z})\\
\textrm{s.t.} \; & \mathbf{z}_j \in\mathbb{PP}_{d_j},\;\forall j \in \mathcal{J},\; 0 \leq  x_i \leq 1,\; \forall i \in \mathcal{I}
\end{aligned}}
\end{equation}}

\vspace{-0.3cm}
\noindent where
\vspace{-0.13cm}
\begin{equation}
\small
\begin{aligned}
&P_{\mu_m}(\mathbf{x},\hat{\mathbf{x}}, \mathbf{z}) \triangleq
 \frac{\mu_m}{2}\sum\limits_{j\in\mathcal{J}}\|\mathbf{P}_j \mathbf{x} - \mathbf{z}_j+\frac{\mathbf{y}_j}{\mu_m}\|^2\\
&+\sum \limits_{i\in\mathcal{I}} \Big(\frac{\mu_m}{2}\big(x_i(\hat x_i-1) + \frac{w_i}{\mu_m}\big)^2
 +\frac{\mu_m}{2}\big(x_i-\hat x_i + \frac{\eta_i}{\mu_m}\big)^2\Big),
\end{aligned}
\end{equation}

\vspace{-0.1cm}
\noindent $\{\mathbf{y}_j\}$, $\{w_i\}$ and $\{\eta_i\}$ denote the dual variables associated with the constraints $\mathbf{P}_j \mathbf{x} = \mathbf{z}_j$, $x_i(\hat x_i-1) = 0$ and $x_i=\hat x_i$, respectively; $\mu_m$ represents the penalty parameter in the $m$-th outer iteration. To this end, we propose to address problem \eqref{AL_problem} by employing the BCD method  in the inner iterations, and update the dual variables  and the penalty
parameter $\{\mu_m\}$ in the outer iterations.

In \eqref{AL_problem}, it can be observed that the primal variables can be divided into three blocks, i.e.,  $\mathbf{x}$, $\hat{\mathbf{x}}$ and $\mathbf{z}$. Therefore, the BSUM iterations for problem \eqref{AL_problem} consists of the following three steps ($k$ denotes the inner iteration index):
\subsubsection{\textbf{Updating $\mathbf{x}^{k+1}$ given $\{\hat{\mathbf{x}}^k,\mathbf{z}^k\}$}}
The $\mathbf{x}$ subproblem is a quadratic optimization problem with a simple constraint that restricts its solution to lie in the interval [0,1], which can be expressed as
\vspace{-0.13cm}
\begin{equation}\label{x_subproblem}
\small
\min \limits_{\mathbf{x}} \;  \mathbf{v}^T \mathbf{x}+ P_{\mu_m}(\mathbf{x},\hat{\mathbf{x}}, \mathbf{z}) \qquad
\textrm{s.t.} \;  0 \leq  x_i \leq 1,\; \forall i \in \mathcal{I}.
\end{equation}

\vspace{-0.2cm}
\noindent We can observe that problem \eqref{x_subproblem} can be naturally decomposed into $N$ subproblems, i.e.,
\vspace{-0.13cm}
\begin{equation} \label{x_subproblem_i}
\small
\min\limits_{x_{i}} \; a_{i,m}^kx_i^2+b_{i,m}^kx_i \qquad
\textrm{s.t.}\; 0 \leq x_i \leq 1,
\end{equation}

\vspace{-0.2cm}
\noindent{where
{ $a_{i,m}^k =\frac{\mu_m}{2}\left(d_i+(\hat{x}_i^k-1)^2+1\right) $}, $b_{i,m}^k = v_i+2\alpha_i+\omega_i(\hat{x}_i^k-1)+\eta_i-\mu_m \hat{x}_i^k$, and $\alpha_i$ denotes the $i$-th element of the vector $\bm{\alpha } = \frac{\mu_m}{2}\sum_{j \in \mathcal{J}} \mathbf{P}_j^T (\frac{\mathbf{y}_j}{\mu_m}-\mathbf{z}_j^k)$.}
By resorting to the first-order optimality condition, the optimal solution of problem \eqref{x_subproblem_i} can be obtained by
\vspace{-0.13cm}
\begin{equation}\label{xup}
\small
x_i^{k+1} = \Pi_{[0,1]}(-0.5b_{i,m}^k/a_{i,m}^k),
\end{equation}

\vspace{-0.2cm}
\noindent where $\Pi_{[0,1]}$ denotes the Euclidean projection operation into the interval $[0,1]$.

\subsubsection{ \textbf{Updating $\mathbf{z}^{k+1}$ given $\{\hat{\mathbf{x}}^k,\mathbf{x}^{k+1}\}$}}
  The optimization problem of $\mathbf{z}_j$ can be expressed as
  \vspace{-0.15cm}
\begin{equation} \label{z_subproblem}
\small
{\min \limits_{\mathbf{z}_j}  \frac{\mu_m}{2}\|\mathbf{P}_j \mathbf{x}^k - \mathbf{z}_j+\frac{\mathbf{y}_j}{\mu_m}\|^2  \qquad
\textrm{s.t.} \;\mathbf{z}_j \in\mathbb{PP}_{d_j}.}
\end{equation}
% As can be seen, the $\mathbf{z}$ subproblem \eqref{z_subproblem} is a quadratic
%optimization problem with a simple constraint that restricts its solution to project onto the parity polytope. Therefore, we can easily obtain its optimal solution by setting the derivative of the objective \eqref{z_subproblem_o} with respect to $\mathbf{z}$ to zero.

\vspace{-0.1cm}
\noindent Similar to the first step, the optimal solution of problem \eqref{z_subproblem} is given by
\vspace{-0.13cm}
\begin{equation}\label{zup}
\small
{\mathbf{z}}_j^{k+1}=\Pi _{\mathbb{PP}_{d_j}}(\mathbf{P}_j \mathbf{x}^k + \frac{\mathbf{y}_j}{\mu_m}),
\end{equation}

\vspace{-0.2cm}
\noindent where $\Pi _{\mathbb{PP}_{d_j}}$ denotes the CPP operation.
\subsubsection{\textbf{Updating $\hat{\mathbf{x}}^{k+1}$ given $\{\mathbf{x}^{k+1},\mathbf{z}^{k+1}\}$}}
The $\hat{\mathbf{x}}$ subproblem can be written as the following unconstrained quadratic optimization problem:
\vspace{-0.13cm}
\begin{equation}
\small
\begin{array}{l}
\min\limits_{\hat{\mathbf{x}}}\; \sum \limits_{i\in \mathcal{I}} \Big(\frac{\mu_m}{2}\big(x_i^k(\hat x_i-1) + \frac{w_i}{\mu_m}\big)^2 +\frac{\mu_m}{2}(x_i^k-\hat x_i + \frac{\eta_i}{\mu_m})^2\Big),
\end{array}
\end{equation}

\vspace{-0.3cm}
\noindent whose optimal solution can be easily obtained by
\vspace{-0.2cm}
\begin{equation}\label{x_up}
\small
{\hat{x}}_i^{k+1}=-\frac{{\mu_m}\left((x_i^k)^2 +1\right)}{4\left((\omega_i-\mu_mx_i^k)x_i^k -(\eta_i+\mu_m x_i^k)\right)}.
\end{equation}
%In summary, all steps in the inner iteration of the problem \eqref{AL_problem} have unique closed-form solutions, and the BSUM algorithm is listed in Algorithm \ref{BSUM}.
%\begin{algorithm}[!h] \footnotesize
%    \caption{BSUM Algorithm for problem \eqref{AL_problem}} \label{BSUM}
%    \begin{algorithmic}[1]
%        \State Obtain $\mathbf{z}^{0}$ and $\hat{\mathbf{x}}^{0}$. Set $k\leftarrow 0$.
%        \Repeat
%        \State Update $\mathbf{x}^{k+1}$, $\mathbf{z}^{k+1}$ and $\hat{\mathbf{x}}^{k+1}$ successively by \eqref{xup}, \eqref{zup} and \eqref{x_up}, respectively.
%        \State $k\leftarrow k+1$.
%        \Until{ some convergence condition is met.} %reach the inner tolerance of accuracy, i.e., $\frac{|P^i-P^{i-1}|}{|P^{i-1}|}\leq \varpi_{\textrm{in}}$ ($P^i$ denotes the
%        \State \textbf{Output:} $\mathbf{x}^{m} \leftarrow \mathbf{x}^{k+1}$, $\mathbf{z}^{m} \leftarrow \mathbf{z}^{k+1}$ and $\hat{\mathbf{x}}^{m} \leftarrow \hat{\mathbf{x}}^{k+1}$,
%    \end{algorithmic}
%\end{algorithm}\vspace{-0.5em}

\vspace{-0.15cm}
Furthermore, the dual variables can be updated by
\vspace{-0.13cm}
\begin{equation} \label{dual_update} \small
\begin{aligned}
\mathbf{y}_j^{m+1} &=
\mathbf{y}_j^m + \mu_m(\mathbf{P}_j \mathbf{x}^m - \mathbf{z}_j^m),\\
w_{i}^{m+1} &= w_{i}^{m} + \mu_m \left(x_i^{m}(\hat x_i^{m}-1)\right),\\
\eta_i^{m+1} &= \eta_i^{m} + \mu_m (x_i^{m}- \hat x_i^{m}).
\end{aligned}
\end{equation}

\vspace{-0.15cm}
\noindent To summarize, the detailed steps of the PDD decoder are listed in Algorithm \ref{PDD}, where $c$ denotes a control parameter that gradually increases the penalty parameter $\mu_m$ by a certain amount during each outer iteration. According to \cite{ShiPDD2017},  the proposed PDD decoder is guaranteed to converge.
\vspace{-0.0cm} \begin{algorithm}[!b]\footnotesize
\setlength{\abovecaptionskip}{-0.5cm}
    \caption{PDD Algorithm for Problem \eqref{LP_problem_equi}} \label{PDD}
    \begin{algorithmic}[1]
        \State Initialize $\mathbf{y}^{0}$, $\{w_{i}\}^{0}$, $\{\eta_i\}^{0}$ and $\mathbf{z}^{0}$ as all-zero vectors. Initialize all elements in $\hat{\mathbf{x}}^{0}$ to $0.5$. Set the initial penalty parameter $\mu_0$ and control parameter $c$. Set $m\leftarrow 0$.
        \Repeat
        \State Obtain $\mathbf{z}^{0}$ and $\hat{\mathbf{x}}^{0}$. Set $k\leftarrow 0$.
        \Repeat
        \State Update $\{\mathbf{x}^{k+1}, \mathbf{z}^{k+1}, \hat{\mathbf{x}}^{k+1}\}$ by \eqref{xup}, \eqref{zup} and \eqref{x_up}. $k\leftarrow k+1$.
        \Until{ some convergence condition is met.} %reach the inner tolerance of accuracy, i.e., $\frac{|P^i-P^{i-1}|}{|P^{i-1}|}\leq \varpi_{\textrm{in}}$ ($P^i$ denotes the
        \State $\mathbf{x}^{m} \leftarrow \mathbf{x}^{k+1}$, $\mathbf{z}^{m} \leftarrow \mathbf{z}^{k+1}$ and $\hat{\mathbf{x}}^{m} \leftarrow \hat{\mathbf{x}}^{k+1}$.

        \State Update the dual variables by \eqref{dual_update} and set $\mu_{m+1} = c \mu_m$.
        \State $\mathbf{z}^{0} \leftarrow \mathbf{z}^{m}$, $\hat{\mathbf{x}}^{0} \leftarrow \hat{\mathbf{x}}^{m}$, $m\leftarrow m+1$.
        \Until{ some convergence condition is met.} %reach the inner tolerance of accuracy, i.e., $\frac{|P^i-P^{i-1}|}{|P^{i-1}|}\leq \varpi_{\textrm{in}}$ ($P^i$ denotes the
    \end{algorithmic}
\end{algorithm}
%\begin{algorithm}[!h] \footnotesize
%    \caption{PDD Algorithm for problem \eqref{LP_problem_equi}} \label{PDD}
%    \begin{algorithmic}[1]
%        \State Initialize $\mathbf{y}^{0}$, $\{w_{i}\}^{0}$, $\{\eta_i\}^{0}$ and $\mathbf{z}^{0}$ as all-zero vectors. Initialize all elements in $\hat{\mathbf{x}}^{0}$ to $0.5$. Set the initial penalty parameter $\mu_0$ and control parameter $c$. Set $m\leftarrow 0$.
%        \Repeat
%        \State Apply Algorithm \ref{BSUM} to obtain the updated $\mathbf{x}^{m}$, $\mathbf{z}^{m}$ and $\hat{\mathbf{x}}^{m}$.
%        \State Update the dual variables by \eqref{dual_update}, set $\mu_{m+1} = c \mu_m$.
%        \State Replace $\mathbf{z}^{0}$ and $\hat{\mathbf{x}}^{0}$ with $\mathbf{z}^{m}$ and $\hat{\mathbf{x}}^{m}$, respectively.
%        \State $m\leftarrow m+1$.
%        \Until{ some convergence condition is met.} %reach the inner tolerance of accuracy, i.e., $\frac{|P^i-P^{i-1}|}{|P^{i-1}|}\leq \varpi_{\textrm{in}}$ ($P^i$ denotes the
%    \end{algorithmic}
%\end{algoithm}
\vspace{-0.25cm}
\section{NCPP algorithm}
The projection $\Pi_{{\mathbb{PP}}_d}(\cdot)$ of a real-valued vector onto the check polytope ${\mathbb{PP}}_d$ in \eqref{zup} is the most time-consuming part in fundamental polytope based decoders, such as the proposed PDD decoder and the ADMM-based decoders in \cite{LiuIT2013} and \cite{7456284}, etc.
In this section, we propose a novel NCPP algorithm which can further reduce the decoding latency of the ICPP algorithm in \cite{Wei2018}. {The main idea of the proposed method is to reduce the number of CPP iterations through a simple three-layer MLP (namely CPP-net) with quantized parameters.} In the following, we first give a brief review of the ICPP algorithm, and then the structure of CPP-net is introduced followed by the proposed NCPP algorithm, and finally we present the detailed process of  training sample generation and loss function design.
%For convenience, we review the details of the iterative CPP algorithm in \cite{Wei2018} as Algorithm \ref{projection}. Algorithm \ref{projection} does not require complex sorting operations, but instead it contains a series of iterations which lie in the iterations of PDD or ADMM. In other words, in each PDD/ADMM iteration, a number of CPP iterations are involved and this could result to considerable decoding latency.
%\subsubsection{CPP-net}

\vspace{-0.45cm}
\subsection{Brief Review of the ICPP Algorithm}
%\begin{algorithm}[!h] \footnotesize
%\setlength\abovedisplayskip{0pt}
%\setlength\belowdisplayskip{-5cm}
%	\caption{Iterative CPP Algorithm} \label{projection}
%	\begin{algorithmic}[1]
%		\State \textbf{Intput:} Vector $\mathbf{v}$ with dimension $d$.
%		\State \textbf{Output:}  Projection $\mathbf{r}=\Pi _{\mathbb{PP}_{d}} (\mathbf{v})$.
%		\State $\theta_i=\textrm{sgn}(v_i-0.5),\;i=1,\cdots,d$.
%		\If{$|\{i:\theta_i = 1\} |$ is even}
%		\State $i^*=\arg\min_i |v_i-0.5|$,
%		\State $\theta_i^*=-\theta_i^*$
%        \EndIf
%		\State $p=|\{i:\theta_i=1\}|-1$
%		\State  $k=0$, $\eta^k=0$,
%		\Repeat
%		\State $\mathbf{v}=\mathbf{v}-\eta^k \bm{\theta}$
%		\State $\mathbf{u}=\Pi_{[0,1]^d} (\mathbf{v})$
%        \State $k=k+1$
%		\State $\eta^k = (\bm{\theta}^T\mathbf{u}-p)/d$
%		\Until{ $\eta^k<\epsilon $} %reach the inner tolerance of accuracy, i.e., $\frac{|P^i-P^{i-1}|}{|P^{i-1}|}\leq \varpi_{\textrm{in}}$ ($P^i$ denotes the
%		\State $\mathbf{r} = \mathbf{u}$
%	\end{algorithmic}
%\end{algorithm}
The ICPP algorithm proposed in \cite{Wei2018} does not require complex sorting operations, however the iterative nature of the algorithm would increase the latency since it lies in each iteration of the proposed PDD decoder and the ADMM-based decoders. { Generally, the ICPP algorithm to obtain $\mathbf{r}=\Pi_{{\mathbb{PP}}_d}(\mathbf{v})$,$\mathbf{v}\in \mathbb{R}^{d_j \times 1}$ works as follows: 1) find the assistant hyperplane $\bm{\theta}$ associated with $\mathbf{v}$, which satisfies ${\bm{\theta}}^T\mathbf{v} = p, p\in \mathbb{R}$  (the value of $p$ can be found by
step 7 of Algorithm 2, which will be introduced later) and determines whether a point in the unit hypercube lies in the check polytope or not, 2) iteratively  derive the difference coefficient $s$ and 3) obtain the projection by
$\mathbf{r}=\Pi_{[0,1]^d}(\mathbf{v}-s\bm{\theta})$.}
%In order to obtain the projection of a real-valued vector $\mathbf{v}\in {\mathbb{R}}^{d_j}$ onto the check polytope, i.e., the CPP $\mathbf{r}=\Pi_{{\mathbb{PP}}_d}(\mathbf{v})$, the work \cite{Wei2018} first finds the assist hypercube $\bm{\theta}$ associated with $\mathbf{v}$, then derives the difference coefficient $s$ and finally obtains the projection by
%$\mathbf{r}=\Pi_{[0,1]^d}(\mathbf{v}-s\bm{\theta})$.
The vector ${\mathbf{v}}'=\mathbf{v}-s\bm{\theta}$ can be interpreted as a shift of vector $\mathbf{v}$ in the direction
orthogonal to the assistant hyperplane $\bm{\theta}$, where the amount
of shift is determined by the value of $s$. In \cite{Wei2018}, an estimate $\hat{s}$ of $s$ was iteratively obtained by $\hat{s}=\sum{\eta^k}$, where ${\eta^k}$ is the incremental projection coefficient and $\eta^k\bm{\theta}$ is how much $\mathbf{v}$ is shifted at the $k$-th iteration. This iterative process terminates when ${\eta^k}$ falls below a certain threshold $\epsilon$.
%where $\eta^o$ satisfies  the following equations:
%\begin{equation}
%\begin{aligned}
%   \mathbf{v}=\mathbf{v}-\eta^o\mathbf{\theta},
%   \mathbf{u} = \Pi_{[0,1]^d} (\mathbf{v}),
%   \bm{\theta}^T\mathbf{u} = p.
%\end{aligned}
%\end{equation}
\vspace{-0.3cm}
\subsection{Structure of CPP-net}
 \begin{figure}[t]
\vspace{-0.1cm}
\setlength{\belowcaptionskip}{-0.6cm}
\renewcommand{\captionfont}{\small}
\centering
\includegraphics[scale=.28]{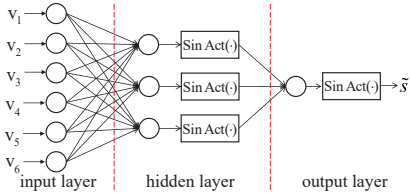}
\caption{Structure of the proposed CPP-net when $d_j=6$.}
\label{NN}
\normalsize
\end{figure}
{ Since the  assistant hyperplane $\bm{\theta}$ is relatively easy to obtain, the main difficulty of
the CPP operation $\mathbf{r}=\Pi_{{\mathbb{PP}}_d}(\mathbf{v})$ lies in the calculation of $s$, which can be viewed as the projection of $\mathbf{v}$ to ${s}$, i.e., $\Pi_{\mathbf{v}\rightarrow{s}}(\mathbf{v})$.} As a result, the CPP operation can be alternatively expressed as
$\mathbf{r} = \Pi_{[0,1]^d}(\mathbf{v}-\Pi_{\mathbf{v}\rightarrow{s}}(\mathbf{v})\bm{\theta})$.
 Motivated by the fact that a trained MLP with  enough neurons can approximate any nonlinear mappings, we introduce a simple three-layer MLP to imitate the  projection $\Pi_{\mathbf{v}\rightarrow s}(v)$  and output an initial estimation of $s$ for the purpose of reducing the residual iteration number. Note that a classical MLP consists of an input layer, an output layer and several hidden layers. Each layer has multiple neurons, and each neuron can execute an activation function on the weighted sum of the
outputs from the preceding layer. The activation function plays an important role in neural networks and when it is non-linear, a two-layer neural network can be proven to be a universal function approximator.
%is propagated through a nonlinear activation function, such as the sigmoid function, the tanh function or the ReLU function.
% Note that a classical MLP consists of an input layer, an output layer, and several hidden layers, and each layer $i$ with $n_i$ inputs and $m_i$ outputs performs
%the mapping ${\mathcal{F}}^{(i)}: {\mathbb{R}}^{n_i} \rightarrow {\mathbb{R}}^{m_i}$  with the weights and biases of the neurons as parameters. There are usually active functions lying  behind each layer to endow the NN with non-linearity, e.g., sigmoid functions, tanh functions or ReLU functions.
%Considering that $\hat{\eta}\ge 0$, we empirically choose $\textrm{sigmoid}(x) = \frac{1}{1+e^{-x}}$  as the active function, whose output is restricted to the interval (0,1). For simplicity of implementation, we propose a new active function called $\textrm{SinAct}(\cdot)$ to approximate $\textrm{tanh}(\cdot)$. $\textrm{SinAct}(\cdot)$ is constructed based on $\sin(\cdot)$, which is usually implemented by ROM on the FPGA platform. Its definition is given as

The proposed CPP-net with $d_j$ inputs consists of three layers, i.e., one input layer with $d_j$ neurons, one hidden layer with $\lceil d_j/2\rceil$ neurons and one output layer with only one neuron.
{ In order to introduce non-linearity into the proposed
network, both  hidden  and  output layers should contain activation functions. Note that the widely-used ReLU activation function is not employed  in the proposed CPP-net
since it will force almost half of the neurons to be silenced (verified by our simulations) and limit the
 learning ability of CPP-net. Instead, we propose a novel activation function constructed based on the $\sin(\cdot)$ function to improve the performance of CPP-net and with low implementation cost.
}
%In order to introduce non-linearity into the proposed network, both the hidden layer and the output layer contain activation functions that are carefully designed for ease of implementation.
We refer to this function as the $\textrm{SinAct}(\cdot)$ and its definition is given by
\vspace{-0.15cm}
\begin{equation}
\small
\textrm{SinAct}(x)=\left\{ \begin{array}{ll}
             \frac{1}{2}\left(\sin(\frac{\pi}{2}x)+1\right),  &-1 \le x \le 1\\
            0,  & x\textless -1\\
            1,  &x \textgreater 1
            \end{array}\right..
\end{equation}

\vspace{-0.15cm}
\noindent For clarity, a simple example of the proposed CPP-net when $d_j=6$ is depicted in Fig. \ref{NN}.

 Let ${\mathbf{y}}^{h}\in{\mathbb{R}}^{3\times1}$ and $\tilde{s}\in \mathbb{R}$ denote the outputs of the hidden and output layers, respectively, then the data flow of CPP-net can be expressed as follows:
%\begin{subequations}
%\begin{align}
%{\mathbf{y}}^{h} = \textrm{SinAct}({\mathbf{W}}_{a}\mathbf{v}+{\mathbf{b}}_a),\\
%\tilde{s} = \textrm{SinAct}({\mathbf{w}}_{b}{\mathbf{y}}^{h}+{\mathbf{b}}_b),
%\end{align}
%\end{subequations}
\vspace{-0.1cm}
\begin{equation}
{\mathbf{y}}^{h} = \textrm{SinAct}({\mathbf{W}}_{a}\mathbf{v}+{\mathbf{b}}_a),\;
\tilde{s} = \textrm{SinAct}({\mathbf{w}}_{b}^T{\mathbf{y}}^{h}+{\mathbf{b}}_b),
\end{equation}

\vspace{-0.1cm}
\noindent where $\bm{\Theta}\triangleq\{{\mathbf{W}}_{a}, {\mathbf{w}}_{b}, {\mathbf{b}}_a, {\mathbf{b}}_b\}$ denote the set of weights and biases, which are the learnable parameters to be trained.  Therefore, the input-output mapping realized by the proposed CPP-net is defined by a chain of functions depending on $\bm{\Theta}$, i.e., $\tilde{s} = \mathcal{F}_{\textrm{CPP-net}}(\mathbf{v};\bm{\Theta})=\textrm{SinAct}({\mathbf{w}}_{b}(\textrm{SinAct}({\mathbf{W}}_{a}\mathbf{v}+{\mathbf{b}}_a))+{\mathbf{b}}_b)$.
%\begin{equation}
%\tilde{s} = \mathcal{F}_{\textrm{CPP-net}}(\mathbf{v};\bm{\Theta})=\textrm{SinAct}({\mathbf{w}}_{b}(\textrm{SinAct}({\mathbf{W}}_{a}\mathbf{v}+{\mathbf{b}}_a))+{\mathbf{b}}_b).
%\end{equation}
%Fig. \ref{NN} depicts an example of CPP-net when $d_j=6$

In order to further reduce the computational complexity  of CPP-net, we propose to quantize the neural weights $\{{\mathbf{W}}_{a},{\mathbf{w}}_{b}\}$ obtained by training to $\{{\mathbf{W}}_{a}^Q,{\mathbf{w}}_{b}^Q\}$, where ${\mathbf{W}}_{a}^Q\in [0,\pm{2^k}]^{\lceil d_j/2\rceil\times d_j},  {\mathbf{w}}_{b}^Q\in[0,\pm{2^k}]^{1\times \lceil d_j/2\rceil},k\in {\mathcal{N}}$ and ${\mathcal{N}}$ represents the set of natural numbers. Let $w^Q$ denote an arbitrary element in $\{{\mathbf{W}}^Q_{a},{\mathbf{w}}_{b}^Q\}$, we can see that the original  multiplication operations involved in the CPP-net can be simplified as follows:
\begin{itemize}
\item If $w^Q = 0$ or $\pm1$, then no multiplication is required.
\item If $w^Q = \pm2^k,k\in{\mathcal{N}}\verb|\|\{0\}$, then the corresponding multiplication operation can be replaced by the binary shifting operation.
\end{itemize}
 Considering that the addition operations are simpler than multiplications, we choose not to quantize the biases $\{{\mathbf{b}}_a, {\mathbf{b}}_b\}$, but instead finetune them with fixed $\{{\mathbf{W}}_{a}^Q,{\mathbf{w}}_{b}^Q\}$. Note that this can compensate the performance loss caused by the quantization of  $\{{\mathbf{W}}_{a},{\mathbf{w}}_{b}\}$, at least to certain extent.

\begin{figure}[t]
\vspace{-0.2cm}
\setlength{\belowcaptionskip}{-0.5cm}
\renewcommand{\captionfont}{\small}
\centering
\includegraphics[scale=.34]{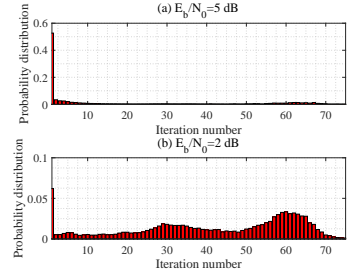}
\caption{Probability distribution of the number of iterations required  by the ICPP algorithm with different $E_b/N_0$ ($\epsilon=10^{-6}$).}
\label{fre}
\normalsize
\end{figure}
\vspace{-0.3cm}
%\subsection{The Details of the Neural CPP algorithm}\vspace{-0.3cm}
\subsection{NCPP algorithm}
 In this subsection, we present the proposed NCPP algorithm, which is shown in Algorithm \ref{projectionNN}. We first decide whether $\mathbf{r}$ can be obtained within only one CPP iteration and if not, we call the CPP-net to obtain an initial estimate $\tilde{s}$ of the difference coefficient $s$. Then, the output of the CPP-net is fed to the subsequent CPP iterations to ensure that an accurate CPP operation can be conducted even when $\tilde{s}$ is far from $s$. Thus, the NCPP algorithm is expected to achieve the same performance as the ICPP algorithm with lower complexity. {Note that the $\textrm{SinAct}(\cdot)$ function can be implemented as a look-up-table and this will not degrade the error correcting performance of the proposed decoder (ensured by steps 12-14 in Algorithm 2). When the number of quantization bits is large enough (e.g., larger than 3), the average iteration number required by the NCPP algorithm is only slightly increased (less than 0.5 in our simulations).}
 \begin{algorithm}[!h] \footnotesize
 \setlength\abovedisplayskip{-10pt}
	\caption{NCPP Algorithm} \label{projectionNN}
	\begin{algorithmic}[1]
		\State \textbf{Intput:} Vector $\mathbf{v}$ with dimension $d_j$.
		\State \textbf{Output:}  $\mathbf{r}=\Pi _{\mathbb{PP}_{d}} (\mathbf{v})$.
		\State $\theta_i=\textrm{sgn}(v_i-0.5),\;i=1,\cdots,d$.
		\If{$|\{i:\theta_i = 1\} |$ is even}
		\State $i^*=\arg\min_i |v_i-0.5|$, $\theta_i^*=-\theta_i^*$.
        \EndIf
		\State $p=|\{i:\theta_i=1\}|-1$, $\mathbf{u}=\Pi_{[0,1]^d} (\mathbf{v})$, $\eta = (\bm{\theta}^T\mathbf{u}-p)/d$.
        \If{$\eta<\epsilon $}
        \State $\mathbf{r} = \mathbf{u}$.
        \Else
        \State $\tilde{s} = \mathcal{F}_{\textrm{CPP-net}}(\mathbf{v})$, $\eta^0 = \tilde{s}$, $k=0$.
		\Repeat
		\State $\mathbf{v}=\mathbf{v}-{\eta^k} \bm{\theta}$, $\mathbf{u}=\Pi_{[0,1]^d} (\mathbf{v})$, $k=k+1$, $\eta^k = (\bm{\theta}^T\mathbf{u}-p)/d$.
		\Until{ $|\eta^k|<\epsilon $} %reach the inner tolerance of accuracy, i.e., $\frac{|P^i-P^{i-1}|}{|P^{i-1}|}\leq \varpi_{\textrm{in}}$ ($P^i$ denotes the
		\State $\mathbf{r} = \mathbf{u}$.
        \EndIf
	\end{algorithmic}
\end{algorithm}
\vspace{-0.5cm}
\subsection{Training Details}
\subsubsection{Training Sample Generation}
 Generally, an MLP is trained to extract the underlying features from  training samples and learn the specific patterns to perform certain tasks, such as classification, clustering and forecasting, etc. Therefore, the performance of the
MLP depends critically on the quality of the training data and in our case, not surprisingly, training with training samples generated under different scenarios will lead to  performance differences   over the same validation set.

Let $({\mathbf{v}}_p,{\hat{s}}_p)_{p=1}^P$ denote the labeled training sample set with  size $P$, where ${\mathbf{v}}_p$ and ${\hat{s}}_p$  represent the $p$-th feature and label, respectively. More specifically, for the considered network, ${\mathbf{v}}_p$ is the input of the CPP operation, which is acquired  by collecting $\mathbf{P}_j \mathbf{x} + \frac{\mathbf{y}_j}{\mu_m}$ in \eqref{zup} when running the PDD decoding algorithm, and the label ${\hat{s}}_p$ is the approximation of $s_p$, which is obtained by running the ICPP algorithm with a predetermined iteration number $K_p$.
%In the proposed CPP-net, there are two main factors during the generation of the training samples that have significant impacts on the network performance, i.e., the iteration number $K_p$ and the training SNR.
Since the proposed network aims to reduce the number of iterations required by the ICPP algorithm, training samples obtained by using different iteration numbers would have a critical impact on the training results. In order to investigate the characteristic of the iteration number, we illustrate its probability distribution when $E_b/N_0$ is set to 2 dB or 5 dB  in Fig. \ref{fre}, where the threshold $\epsilon$ is fixed to $10^{-6}$.\footnote{For the detailed simulation setup, please refer to Fig. 4 (a). Note that $E_b/N_0$=2 dB corresponds to the low SNR scenario, while $E_b/N_0$=5 dB denotes the high SNR scenario.} we can observe that for both cases, the proportion of $K_p=1$ (i.e., the ICPP algorithm converges within only one iteration) is larger than the others. Since employing CPP-net is unnecessary when $K_p=1$, the training samples obtained when $K_p=1$ are useless for network training and these instances should not be included in the training sample set. In addition, considering that high noise levels would prevent the proposed network from learning the underlying mapping mechanism, the training samples  with $K_p\ge 2$ are collected under a relatively high $E_b/N_0$ ($E_b/N_0$=5 dB is used in our simulations).

\subsubsection{Loss Function}
Loss function is used
to measure the differences between the network output and the true label, and the performance of the network is heavily dependent on it.
In general, the loss function should be carefully defined according to  the specific learning task.
In the following, we first investigate the convergence property of the proposed NCPP algorithm, based on which we present a novel loss function that is able to accelerate the learning process.

\begin{figure}[t]
\vspace{-0.2cm}
\setlength{\belowcaptionskip}{-0.5cm}
\renewcommand{\captionfont}{\small}
\centering
\includegraphics[scale=.36]{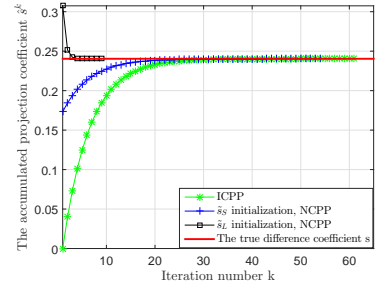}
\caption{Convergence behavior of the NCPP algorithm with different values of $\tilde{s}$.}
\label{ls}
\normalsize
\end{figure}

In Fig. \ref{ls}, we illustrate the typical convergence behaviors of the proposed NCPP algorithm with different values of $\tilde{s}$, where $s$ denotes the true difference coefficient, $\tilde{s}_L$ and $\tilde{s}_S$ denote two initial estimates of $s$ with $\tilde{s}_L\textgreater s,\; \tilde{s}_S\textless s$ and $\tilde{s}_L-s=s-\tilde{s}_S\textgreater \epsilon$. { Note that the CPP-net can be viewed as a non-linear projector which is
able to output an approximate value of  the difference coefficient from the input $\mathbf{v}$, therefore, it is able to provide a good initial  point for the ICPP algorithm.}
The accumulated projection coefficient up to iteration $k$, i.e., ${\hat{s}}^{k}=\sum_{i=0}^k\eta^i$,  is regarded as the performance metric. Note that the ICPP algorithm can be viewed as a special case of the proposed NCPP algorithm with $\tilde{s}=0$.  we can observe that different values of $\tilde{s}$ lead to different numbers of iterations with the same $\epsilon$ even when $|\tilde{s}_L-s| = |\tilde{s}_S-s|$, and taking $\tilde{s}_L$ as the initial point results in a smaller iteration number.
%Let $\eta$ represent the cumulative value of $\eta^k$ with iterations, and our evaluation of convergence is based on the change of $\eta$ when iteration increases.
%The  convergency process of $\eta$ can be divided into two scenarios \cite{Wei2018}, as shown in Fig. \ref{conv}. In Scenario (a), $\eta$ can converge to ${\eta}^o$ with a small number of iterations, while
%in Scenario (b), $\eta$ infinitely approaches ${\eta}^o$ as the iteration number increases, but it can't reach the exact value, so the required iteration number $K$ could be very large. Thus, for Scenario (a), what we can do for the reduction of iteration numbers is quite limited. Hence, the CPP-net is mainly trained for reducing $K$ in Scenario (b).
%The basic idea is to make $\hat{\eta}$ approach $\eta^o$ as close as possible by minimizing $||\eta_o-\hat{\eta}||_2^2$.
%Furthermore, it is important to note that a large approximate value of $\eta_o$ and a small one will have a different impact on iteration number. Specifically, let $\hat{\eta}_L$ and $\hat{\eta}_S$ denote two estimates of $\eta^o$ obtained by CPP-net with $\hat{\eta}_L\textgreater \eta^o,\; \hat{\eta}_S\textless \eta^o$ and $\hat{\eta}_L-\eta^o=\eta^o-\hat{\eta}_S\textgreater \epsilon$. Fig. \ref{ls} shows the convergence process of  $\eta$ with different starting point $\hat{\eta}$. Since the convergence velocity becomes slow with the increasing of iterations, multiple iterations are still required with  ${\hat{\eta}}_S$ even though it is close to $\eta^o$. On the contrary, with a little larger starting point $\hat{\eta}_L$, $\eta$ can converge to $\eta^o$ within only a few iterations. Therefore, the output of CPP-net which is a little larger than $\eta^o$ is preferred.
Based on this observation, we design the loss function as
%\begin{equation}
%\mathcal{L}_{\textrm{CPP-net}}(\bm{\Theta}) = \frac{1}{P}\sum_{p=1}^{P}({\hat{s}}_p-\tilde{s}_p+\kappa||{\hat{s}}_p-\tilde{s}_p||_2^2),
%\end{equation}
$\mathcal{L}_{\textrm{CPP-net}}(\bm{\Theta}) = \frac{1}{P}\sum_{p=1}^{P}({\hat{s}}_p-\tilde{s}_p+\kappa||{\hat{s}}_p-\tilde{s}_p||_2^2)$,
where the coefficient $\kappa$ is a weighting factor (hyperparameter) which needs to be predefined before training. We can see that the proposed loss function consists of two terms, i.e., $\hat{s}_p-\tilde{s}_p$ and  $||\hat{s}_p-\tilde{s}_p||_2^2$, $\hat{s}_p-\tilde{s}_p$ is designed such that a larger initial estimate of $s$ is preferred and $||\hat{s}_p-\tilde{s}_p||_2^2$ is used to minimize the difference between the network output and the label.
%\begin{figure*}[t]
%\begin{subequations}\label{W12}
%\begin{align}
%&\{{\mathbf{W}}_{12}^Q,{\mathbf{W}}_{23}^Q\}_{{\mathcal{C}}_1}=\left\{\left[\begin{matrix} 0&0&0&-2&0&0 \\ 0&0&0&-2&0&0 \\1&1&-1 &0&0&0 \end{matrix}\right]^T,[-1,1,0]^T\right\}\\
%&\{{\mathbf{W}}_{12}^Q,{\mathbf{W}}_{23}^Q\}_{{\mathcal{C}}_3,6}=\left\{\left[\begin{matrix} 0&-2&0&0&0&0 \\ 0&1&0&0&0&0 \\0&0&0 &0&0&2 \end{matrix}\right]^T,[2,2,-1]^T\right\}\\
%&\{{\mathbf{W}}_{12}^Q,{\mathbf{W}}_{23}^Q\}_{{\mathcal{C}}_3,7}=\left\{\left[\begin{matrix} -2&-2&0&-1&0&0&0 \\ 0&0&0&0&-1&0&0 \\0&-1&0 &0&0&0&0\\0 &0&0&0&-1&0&0 \end{matrix}\right]^T,[0,2,2,-2]^T\right\}
%\end{align}
%\end{subequations}
%\end{figure*}
\vspace{-0.2cm}
\section{Simulation Result}
In this section, computer simulations are carried out to evaluate the error-correcting performance of the proposed PDD decoder and the decoding latency  of the NCPP algorithm. The proposed network is implemented in Python using the TensorFlow library with the Adam optimizer \cite{ADAM}. In the simulations, we focus on additive white Gaussian noise channel with binary phase shift keying (BPSK) modulation. { The considered binary linear codes are (96, 48)  MacKay 96.33.964 LDPC code $\mathcal{C}_1$, (575, 288) IEEE 802.16e LDPC code $\mathcal{C}_2$  and (2640, 1320) Margulis code $\mathcal{C}_3$ \cite{channelcodes}.}
%\begin{figure*}[t]
%\begin{subequations}\label{W12}
%\begin{align}
%&\{{\mathbf{W}}_{12}^Q,{\mathbf{W}}_{23}^Q\}_{{\mathcal{C}}_1}=\left\{\left[\begin{matrix} 0&0&0&-2&0&0 \\ 0&0&0&-2&0&0 \\1&1&-1 &0&0&0 \end{matrix}\right]^T,[-1,1,0]^T\right\}\label{w1}\\
%&\{{\mathbf{W}}_{12}^Q,{\mathbf{W}}_{23}^Q\}_{{\mathcal{C}}_2,6}=\left\{\left[\begin{matrix} 0&-2&0&0&0&0 \\ 0&1&0&0&0&0 \\0&0&0 &0&0&2 \end{matrix}\right]^T,[2,2,-1]^T\right\}\\
%&\{{\mathbf{W}}_{12}^Q,{\mathbf{W}}_{23}^Q\}_{{\mathcal{C}}_2,7}=\left\{\left[\begin{matrix} -2&-2&0&-1&0&0&0 \\ 0&0&0&0&-1&0&0 \\0&-1&0 &0&0&0&0\\0 &0&0&0&-1&0&0 \end{matrix}\right]^T,[0,2,2,-2]^T\right\}
%\end{align}
%\end{subequations}
%\end{figure*}
During the training process, we collect $10^5$ training samples and $10^4$ validation samples with $E_b/N_0=5$ dB, $E_b/N_0=4.5$ dB and $E_b/N_0=3$ dB for ${\mathcal{C}}_1$, ${\mathcal{C}}_2$ and ${\mathcal{C}}_3$ codes. The learning rate  and the balance coefficient $\kappa$ are set to $10^{-4}$ and 4.

\begin{figure}[t]
\vspace{-0.2cm}
\setlength{\belowcaptionskip}{-0.5cm}
\renewcommand{\captionfont}{\small}
\centering
\includegraphics[scale=.42]{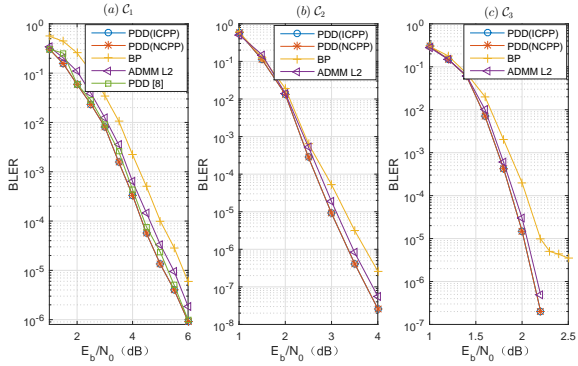}
\caption{ BLER performance comparison of $\mathcal{C}_1$, $\mathcal{C}_2$ and ${\mathcal{C}}_3$ codes.}
\label{BLER}
\normalsize
\end{figure}

{ We first compare the BLER performance of  the proposed PDD decoder, the BP decoder (sum-product), the ADMM $\ell_2$  decoder in \cite{7456284} and the PDD decoder in \cite{8691508},  as shown in Fig. \ref{BLER}}.\footnote{{ Note that for the considered codes, we have tested the  ADMM penalized decoders with many other penalty functions, and we finally chose the ADMM $\ell_2$ decoder in terms of BLER performance.}} In all the curves, we collect at least 100 block errors for all data points. It can be observed that our proposed PDD decoder shows better BLER performance at both low and high SNR regions for ${\mathcal{C}}_1$ code. {For longer LDPC codes, i.e., the ${\mathcal{C}}_2$ and ${\mathcal{C}}_3$ codes, the proposed PDD decoder achieves a similar performance as the other counterparts when the SNR is low and outperforms them when $E_b/N_0\ge 2$ dB and $E_b/N_0\ge 1.4$ dB for ${\mathcal{C}}_2$ and ${\mathcal{C}}_3$ codes, respectively. Specifically, 0.3 dB, 0.1 dB and 0.08 dB performance gains over the ADMM $\ell_2$ decoder can be achieved at BLER=$10^{-4}$ for ${\mathcal{C}}_1$, ${\mathcal{C}}_2$ and ${\mathcal{C}}_3$ codes, respectively.} { Besides, although the proposed PDD decoder achieves a  similar BLER performance as that in \cite{8691508}, it requires less auxiliary variables and thus potentially leads to lower complexity.}

{ Then, in TABLE \ref{Tab1}, we provide the iteration numbers required by the ICPP algorithm \cite{Wei2018} and Algorithm \ref{projectionNN} when decoding ${\mathcal{C}}_1$ ($d=6$), ${\mathcal{C}}_2$ ($d=6$ or 7) codes and (128, 64) CCSDS code ${\mathcal{C}}_4$ ($d=8$) \cite{channelcodes} with $E_b/N_0 = 3$ dB and $\epsilon=10^{-6}$.
 Since the average (\emph{Ave}) and worst case (\emph{Wor}) iteration numbers required by the CPP operation both affect the decoding latency and  throughput, we choose them as the performance metrics.
It can be seen from TABLE \ref{Tab1} that the proposed CPP-net can reduce both the average and worst case iteration numbers and in particular, the average iteration number is reduced by nearly half.}
%\begin{table}[t]
%\vspace{-1cm}
%\setlength{\abovecaptionskip}{-0.0cm}
%\setlength{\belowcaptionskip}{-0.0cm}
%\caption{Iteration number ($E_b/N_0=3 \textrm{dB}$, $\epsilon=10^{-6}$).}
%\begin{center}
%\begin{tabular}{|c|c|c|c|c|}
%\hline
%\multirow{2}*{} & \multicolumn{2}{c|}{${\mathcal{C}}_1$} & \multicolumn{2}{c|}{${\mathcal{C}}_2$}\\
%\cline{2-5}
%&{IterM} & {IterM1} & {IterM} & {IterM1}\\
%\hline
%Iterative CPP & 20.3675 &34.3724 &28.7205 &42.8809\\
%\hline
%Neural CPP & 11.0653 &18.3436 &15.7024 &23.2127\\
%\hline
%\end{tabular}\vspace{-0.6cm}
%\label{Tab1}
%\end{center}
%\end{table}
%Let $\{{\mathbf{W}}_{a}^Q,{\mathbf{W}}_{b}^Q\}_{{\mathcal{C}}_1}$, $\{{\mathbf{W}}_{a}^Q,{\mathbf{W}}_{b}^Q\}_{{\mathcal{C}}_2,6}$ and $\{{\mathbf{W}}_{a}^Q,{\mathbf{W}}_{b}^Q\}_{{\mathcal{C}}_2,7}$ denote the trained $\{{\mathbf{W}}_{a}^Q,{\mathbf{W}}_{b}^Q\}$ for ${\mathcal{C}}_1$, ${\mathcal{C}}_2$ with 6-dimension $\mathbf{v}$ and ${\mathcal{C}}_2$ with 7-dimension $\mathbf{v}$, respectively. As listed in (\ref{W12}), there are multiple zero elements in these trained weights, contributing to low computational resumption of CPP-net.

Finally, we provide a computational complexity
analysis of the ICPP and NCPP algorithms, which is based on the numbers of multiplications (Muls) and additions (Adds) required by the CPP operation. For simplicity, we take ${\mathcal{C}}_1$ code as an example and the analysis for $C_2$ code can be similarly conducted. Note that the complexity of one CPP iteration (step 13 in Algorithm \ref{projectionNN}) involves:
1) updating $\mathbf{v}$, which requires $d$ Muls and $d$ Adds;
%2) calculating $\eta$, where the division operation that $(\bm{\theta}^T\mathbf{u}-p)$ is divided by $d$ can be viewed as a multiplication operation that $(\bm{\theta}^T\mathbf{u}-p)$ is multiplied by $1/d$,  which requires a complexity of $2d$ Mul + $d$ Add.
2) calculating $\eta$ needs $2d$ Muls and $d$ Adds.
 For ${\mathcal{C}}_1$ code, we list the quantized parameters $\{{\mathbf{W}}_{a}^Q,{\mathbf{w}}_{b}^Q\}$ of the CPP-net as follows:
\begin{equation}
\small
{\mathbf{W}}_{a}^Q=\left[\begin{matrix} 0&0&0&-2&0&0 \\ 0&0&0&-2&0&0 \\1&1&-1 &0&0&0 \end{matrix}\right]^T, {\mathbf{w}}_{b}^Q=[-1,1,0]^T.\label{w1}
\end{equation}
{Therefore, the complexity of one forward pass of the CPP-net can be expressed as $2$ Muls and $7$ Adds. Based on the average iteration number in TABLE \ref{Tab1}, the average numbers of Adds and Muls required by the ICPP and NCPP algorithms are listed as TABLE II.}
 %366.6150 Muls +  244.4100 Adds and 206.174 Muls + 139.7836 Adds, respectively.
Given the fact that the CPP operations are needed in each iteration of the proposed PDD decoder or the ADMM $\ell_2$ decoder, employing Algorithm \ref{projectionNN} is able to reduce the computational complexity and decoding latency of these decoders significantly.
%\begin{table}[t]
%\vspace{-0.5cm}
%\setlength{\abovecaptionskip}{-0.0cm}
%\setlength{\belowcaptionskip}{-0.0cm}
%\caption{Iteration number ($E_b/N_0=3 \textrm{dB}$, $\epsilon=10^{-6}$).}
%\begin{center}
%\begin{tabular}{|c|c|c|c|c|}
%\hline
%\multirow{2}*{} & \multicolumn{2}{c|}{${\mathcal{C}}_1$} & \multicolumn{2}{c|}{${\mathcal{C}}_2$}\\
%\cline{2-5}
%&{IterM} & {IterM1} & {IterM} & {IterM1}\\
%\hline
%Iterative CPP & 20.3675 &34.3724 &28.7205 &42.8809\\
%\hline
%Neural CPP & 11.0653 &18.3436 &15.7024 &23.2127\\
%\hline
%\end{tabular}\vspace{-0.6cm}
%\label{Tab1}
%\end{center}
%\end{table}
%\begin{table}[t]
%\vspace{-0.3cm}
%\setlength{\abovecaptionskip}{-0.0cm}
%\setlength{\belowcaptionskip}{-0.0cm}
%\caption{Iteration number ($E_b/N_0=3 \textrm{dB}$, $\epsilon=10^{-6}$).}
%\begin{center}
%\begin{tabular}{|c|c|c|}
%\hline
%&{IterM} of ${\mathcal{C}}_1$ & {IterM} of ${\mathcal{C}}_2$ \\
%\hline
%Iterative CPP & 20.3675  &28.7205 \\
%\hline
%Neural CPP & 11.0653  &15.7024 \\
%\hline
%\end{tabular}\vspace{-0.6cm}
%\label{Tab1}
%\end{center}
%\end{table}

%\begin{table}[t]
%\vspace{-0.0cm}
%\small
%\setlength{\abovecaptionskip}{-0.0cm}
%\setlength{\belowcaptionskip}{-0.0cm}
%\caption{Iteration number ($E_b/N_0=3$ dB, $\epsilon=10^{-6}$).}
%\begin{center}
%\begin{tabular}{|c|c|c|c|}
%\hline
%&${\mathcal{C}}_1$ ($d=6$)&${\mathcal{C}}_2$ ($d=6/7$) &  ${\mathcal{C}}_4$ ($d=8$)\\
%\hline
%ICPP & 20.3675  &28.7205 &  24.7334\\
%\hline
%NCPP & 11.0653  &15.7024 &  13.5614 \\
%\hline
%\end{tabular}\vspace{-0.3cm}
%\label{Tab1}
%\end{center}
%\end{table}

\begin{table}[t]
\vspace{-0.0cm}
\small

\setlength{\abovecaptionskip}{-0.0cm}
\setlength{\belowcaptionskip}{-0.0cm}
\caption{  Iteration number comparison.}
\begin{center}
\begin{tabular}{|c|c|c|c|c|c|c|}
\hline
\multirow{2}{*} {} & \multicolumn{2}{c|}{${\mathcal{C}}_1$ ($d=6$)} & \multicolumn{2}{c|}{${\mathcal{C}}_2$ ($d=6/7$)}& \multicolumn{2}{c|}{${\mathcal{C}}_4$ ($d=8$)}\\
\cline{2-7}
  & \emph{Ave} & \emph{Wor} & \emph{Ave} & \emph{Wor} & \emph{Ave} & \emph{Wor}\\
\hline
ICPP & 20.3675 & 72  &28.7205 & 89 & 24.7334 & 79\\
\hline
NCPP & 11.0653 & 61 &15.7024 & 73 & 13.5614 & 68 \\
\hline
\end{tabular}\vspace{-0.3cm}
\label{Tab1}
\end{center}
\end{table}
%\begin{table}[t]
%\vspace{0.1cm}
%\setlength{\abovecaptionskip}{-0.0cm}
%\setlength{\belowcaptionskip}{-0.0cm}
%\caption{ Computation complexity comparison for ${\mathcal{C}}_1$ ($E_b/N_0=3 \textrm{dB}$, $\epsilon=10^{-6}$).}
%\begin{center}
%\begin{tabular}{|c|c|c|}
%\hline
%&  Muls &  Adds  \\
%\hline
% Iterative CPP &   366.6150  & 244.4100 \\
%\hline
% Neural CPP &   206.174  &   139.7836  \\
%\hline
%\end{tabular}\vspace{-0.7cm}
%\label{Tab1}
%\end{center}
%\end{table}
\begin{table}[t]
\vspace{-0.0cm}
\setlength{\abovecaptionskip}{-0.1cm}
\setlength{\belowcaptionskip}{-0.0cm}
\footnotesize
\caption{Computation complexity comparison.}
\begin{center}
\begin{tabular}{|c|c|c|c|c|c|c|}
\hline
\multirow{2}{*} {} & \multicolumn{2}{c|}{${\mathcal{C}}_1$} & \multicolumn{2}{c|}{${\mathcal{C}}_2$}& \multicolumn{2}{c|}{${\mathcal{C}}_4$}\\
\cline{2-7}
  & Muls & Adds & Muls & Adds & Muls & Adds\\
\hline
ICPP &  366.61  &244.41 & 560.05 & 373.37 & 593.61 & 395.73\\
\hline
NCPP & 201.17  &  139.78 & 308.19 & 212.63 & 328.47 & 232.98 \\
\hline
\end{tabular}\vspace{-0.6cm}
\label{Tab2}
\end{center}
\end{table}
\vspace{-0.2cm}
\section{Conclusion}
In this work, we presented a novel PDD decoder with for binary linear codes.
% We transformed the original ML decoding problem into its equivalent form by introducing auxiliary variables, then we employed PDD framework to solve the resulting problem with coupling constraints. Furthermore, inspired by the truth that multi-layer NN can theoretically map any non-linear relation, we introduced a specially designed simple three-layer NN into the iterative CPP algorithm to decrease the computational complexity by reducing the number of the iterations.
We showed that other than the minimum polytope based LP problem, the PDD framework can also be utilized to address the  fundamental polytope based LP decoding problem. Furthermore, a NCPP algorithm was proposed to reduce the iteration number required by the ICPP algorithm, and it is applicable to all ADMM or PDD based decoders that involve the CPP operations.
Simulation results demonstrated the superior performance of the proposed PDD decoder and the effectiveness of the NCPP algorithm for complexity and latency reduction.

% or


%\appendices
%% you can choose not to have a title for an appendix
%% if you want by leaving the argument blank
%\section{}
%% use section* for acknowledgment
%\section*{Acknowledgment}

\ifCLASSOPTIONcaptionsoff
  \newpage
\fi



%\bibliographystyle{IEEEtran}
%\begin{thebibliography}{1}
%\end{thebibliography}
\vspace{-0.5em}
\bibliography{ADMM}
\bibliographystyle{IEEETran}


%\end{IEEEbiography}


%\newpage



%\vfill


\end{document}